\documentclass[%
 reprint,
 amsmath,amssymb,
 aps,
prb
,superscriptaddress
]{revtex4-1}

\usepackage{subfigure}
\usepackage{graphicx}
\usepackage{tabularx}
\usepackage{dcolumn}
\usepackage{bm}

\begin{document}

\preprint{APS/123-QED}

\title{Theory of the Spin Seebeck Effect at a Topological-Insulator/Ferromagnetic-Insulator
Interface}


\author{Nobuyuki Okuma}\email{okuma@hosi.phys.s.u-tokyo.ac.jp}
\affiliation{Department of Physics, University of Tokyo, Hongo 7-3-1, 113-0033, Japan}
\author{Massoud Ramezani Masir}
\affiliation{Department of Physics, University of Texas at Austin, Austin, TX 78712, USA}
\author{Allan H. MacDonald}
\affiliation{Department of Physics, University of Texas at Austin, Austin, TX 78712, USA}

%
\date{\today}

\begin{abstract} 
The spin-Seebeck effect refers to voltage signals induced in metals by thermally driven
spin currents in adjacent magnetic systems.  We present a theory of the spin-Seebeck 
signal in the case where the conductor that supports the voltage signal is the topologically protected
two-dimensional surface-state system at the interface between a ferromagnetic insulator (FI) and a 
topological insulator (TI).  Our theory uses a Dirac model for the TI surface-states 
and assumes Heisenberg exchange coupling between the TI quasiparticles and the FI magnetization.
The spin-Seebeck voltage is induced by the TI surface states scattering off the 
nonequilibrium magnon population at the surface of the semi-infinite thermally driven FI.
Our theory is readily generalized to spin-Seebeck voltages 
in any two-dimensional conductor that is exchange-coupled to the surface of a FI.
Surface-state carrier-density-dependent signal strengths calculated 
using Bi$_2$Te$_3$ and yttrium iron garnet  
material parameters are consistent with recent experiments.
\begin{description}
\item[PACS numbers]
72.20.Pa, 72.25.Mk, 73.20.-r, 75.76.+j
\end{description}
\end{abstract}

\maketitle

\section{INTRODUCTION}
The spin-Seebeck effect (SSE),\cite{uchida2008,uchida2010,longuchida,dqu,kikkawa} in which 
the spin-current response to a temperature gradient in a ferromagnet gives rise to a voltage signal in an adjacent metal,
has emerged as a central issue of spin caloritronics.\cite{caloritronics,caloritronics2}
In the case of bilayers\cite{uchida2010,longuchida,kikkawa,dqu} 
formed by an insulating ferromagnet and a nonmagnetic metal,
for example Pt and yttrium iron garnet (YIG), 
the signal is interpreted \cite{uchida2010,longuchida,kikkawa} 
as an inverse spin Hall effect (ISHE) voltage associated with
conversion between magnon and electron spin currents 
at the normal-metal/ferromagnetic-insulator (FI) interface.  
The FI is often modeled as a magnon gas\cite{adachi2,slzhang1,slzhang2,bauer}, 
with classical dynamics described by the stochastic Landau-Lifshitz-Gilbert equation.\cite{hoffman}
Adachi $et$ $al.$ explained the SSE by a quantum theory with a  
temperature difference between electrons in the normal metal and magnons in the FI\cite{adachi2}.
Semiclassical theories rest on a description of the conversion of the magnon spin current generated 
by a thermal gradient to the electron spin current\cite{slzhang1,slzhang2,bauer} at the FI-metal interface.

\begin{figure}[]
\begin{center}
\includegraphics[width=8cm,angle=0,clip]{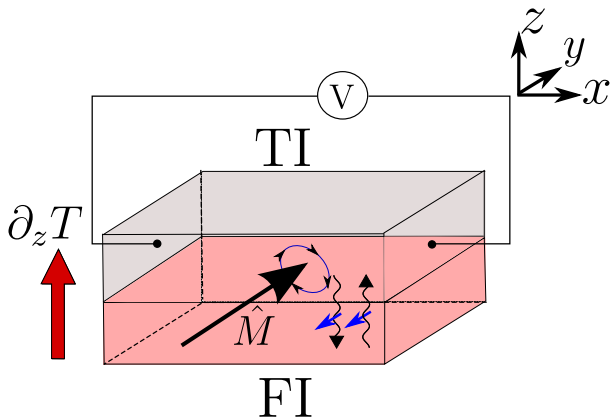}
\caption{Schematic illustration of the spin-Seebeck effect in a topological-insulator/ferromagnetic-insulator (FI) bilayer.
$\hat{M}\parallel \hat{y}$ indicates the FI ground-state magnetization direction. 
A finite voltage in the $x$ direction is generated by a vertical thermal gradient $\partial_zT$. }
\label{fig1}
\end{center}
\end{figure}

Recently\cite{jiang} a spin-Seebeck signal has been measured at the interface between the topological insulator 
(TI) Sb-doped Bi$_2$Te$_3$ and YIG (see Fig. \ref{fig1}). 
This experiment provides an example of a SSE voltage signal induced 
in a two-dimensional conductor that is coupled to the surface of a FI.
Since the bulk of the TI does not support a spin current, it is clear that 
the SSE voltage signal generation mechanism must differ from the ISHE mechanism thought to act in a   
FI/non-magnetic-metal bilayer.   SSE experiments are 
normally interpreted in terms of momentum-averaged quantities such as the  
total spin current\cite{slzhang1,slzhang2,bauer}.
Because the TI surface states are coupled to the FI via exchange interactions,
the signal must\cite{hasan,xlq,jiang} originate from TI surface-state quasiparticles scattering
off the nonequilibrium magnon population at the FI surface.  As we show, the spin-Seebeck voltage then 
depends on the full momentum nonequilibrium distribution of magnons evaluated 
at the FI surface, and not just on the nonequilibrium magnon density. 

In this paper, we present a theory for the SSE observed in a TI/FI bilayer that is 
based on the semiclassical transport theory applied to the bulk of the FI and to the 
TI/FI interface.  We show that the nonequilibrium magnon population at the FI 
surface drives a charge current at the TI surface.  
By analyzing the magnon and electron Boltzmann equations, which well describe nonequilibrium transport under the static electromagnetic field and thermal gradient, we obtain 
an expression of the electric field induced at the TI surface under open 
circuit conditions.  
Our theory can be easily generalized to any two-dimensional conductor at a surface of a magnetic 
material, e.g., to graphene on YIG.

The paper is organized as follows.
In Sec. \ref{magsec}, we solve the steady-state magnon Boltzmann equation of a semi-infinite
FI in the presence of a thermal gradient oriented perpendicular to the surface, assuming specular 
scattering of magnons, and extract results for the magnon distribution at the surface.  
In Sec. \ref{modelsec}, we consider the spin-momentum-locked Dirac electrons at the 
TI surface and account for exchange coupling to the FI.
Using the nonequilibrium magnon distribution function obtained in Sec. \ref{magsec}, we evaluate the
nonzero net rate of transitions in the TI surface-state system induced by  
scattering off the FI's nonequilibrium magnon population
and use it to obtain an expression for the electric field
induced under open circuit conditions.  
In Sec. \ref{numerics}, we estimate the typical size of the SSE using materials parameters appropriate for Bi$_2$Te$_3$ and YIG
and compare our results with experimental data.
Some related effects in other hybrid materials 
are discussed in Sec. \ref{discussion}.

\section{Magnon distribution at the surface of a ferromagnetic insulator with a perpendicular 
temperature gradient\label{magsec}}
In this section, we use the magnon Boltzmann equation with a specular-reflection boundary condition
to calculate the magnon distribution function at the surface of a FI with a perpendicular 
temperature gradient.  Henceforth, we set $\hbar=k_B=1$.
\begin{figure}[]
\begin{center}
\includegraphics[width=7cm,angle=0,clip]{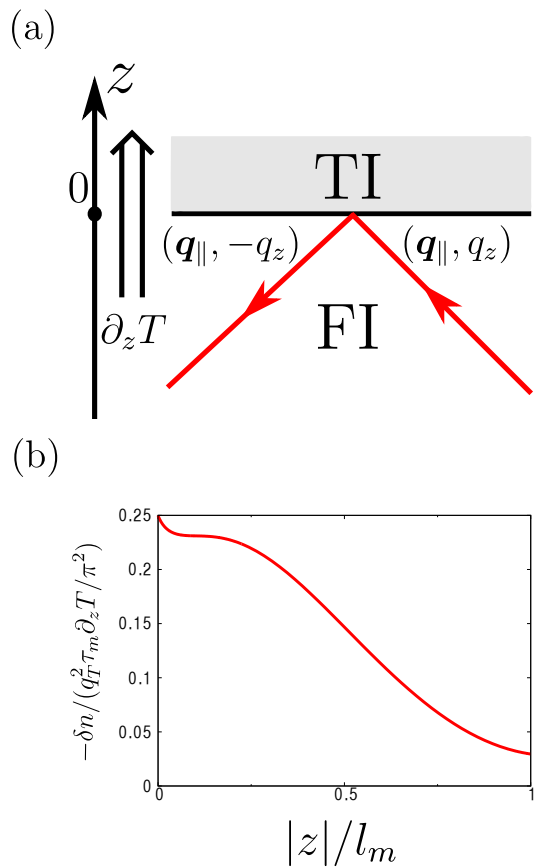}
\caption{(a) Schematic illustration of specular reflection of magnons at the 
FI/TI interface. (b) Total excess magnon population $\delta n$ as a function of
the ratio of the distance from the interface $|z|$ to the magnon mean free-path $l_m$.}
\label{fig2}
\end{center}
\end{figure}
\subsection{Model of a ferromagnetic insulator}
We consider a FI with magnetization in the $y$ direction as illustrated in Fig. \ref{fig1}. 
The low-energy spin excitations of the FI can be described by magnon creation and annihilation 
operators [$a(\bm{x}), a^{\dagger}(\bm{x}))$].  In the case of a quantum spin model with 
spin $S_0$ degrees of freedom on each lattice site, creation 
and annihilation operators can be introduced by the Holstein-Primakoff transformation:
$S^y= S_0-a^{\dagger}a$, $S^z+iS^x\simeq\sqrt{2S_0}a$, and $S^z-iS^x\simeq\sqrt{2S_0}a^\dagger$, where 
$S^i$ is a spin-operator component.
In the following, we neglect magnetic anisotropy.  
The Holstein-Primakoff transformation then 
leads to a three-dimensional magnon gas with
isotropic quadratic dispersion $\omega_{\bm{q}}=D|\bm{q}|^2$
at long wavelengths.  Here $D$ is the spin stiffness, and $\bm{q}$ is the three-dimensional magnon momentum. 
In terms of magnon operators, the low-energy effective Hamiltonian for the FI is given by
\begin{align}
H_{m}= V \int \frac{d^3q}{(2\pi)^3}\omega_{\bm{q}}a^\dagger_{\bm{q}}a_{\bm{q}},
\end{align}
where $V$ is volume.  At low energies, this magnon-gas model applies equally well to
ferrimagnetic insulators like YIG with a net magnetization due to incomplete 
cancellation between antiferromagnetically aligned spins.

\subsection{Magnon Boltzmann equation}
We now consider the magnon Boltzmann equation in the presence of a thermal gradient:
\begin{align}
\frac{\partial n_{\bm{q}}}{\partial t}+v_{q_z}\partial_zn_{\bm{q}}=\left.\frac{\partial n_{\bm{q}}}{\partial t}\right|_{scatt},\label{magboltz}
\end{align}
where $n_{\bm{q}}$ is the momentum-dependent magnon distribution function, $v_{q_z}=\partial_{q_z} \omega_{\bm{q}}$ is the magnon velocity, and the right-hand-side term is the scattering term.
In the following, we adopt the relaxation-time approximation in which the scattering term is given by
\begin{align}
\left.\frac{\partial n_{\bm{q}}}{\partial t}\right|_{scatt}=-\frac{n_{\bm{q}}-n^{(0)}_{\bm{q}}(T(z))}{\tau_m},\label{magrelaxterm}
\end{align}
where $n^{(0)}_{\bm{q}}=(\exp(\omega_{\bm{q}}/T(x))-1)^{-1}$ is the Bose distribution function with local temperature $T(z)$, and $\tau_m$ is a magnon relaxation time.
The validity of the relaxation-time approximation for the 
scattering term ($\ref{magrelaxterm}$) is discussed in Sec. \ref{discussion}.

For linear response to a temperature gradient,
the Boltzmann equation becomes
\begin{align}
v_{q_z}\left[\partial_z(\delta n_{\bm{q}})+\partial_zT\frac{\partial n^{(0)}_{\bm{q}}}{\partial T}\right]=-\frac{\delta n_{\bm{q}}}{\tau_m},\label{steadymagboltz}
\end{align}
where $\delta n_{\bm{q}}\equiv n_{\bm{q}}-n^{(0)}_{\bm{q}}$ is the magnon distribution response.
In the following, we solve Eq. ($\ref{steadymagboltz}$) assuming specular reflection of magnons
at the surface of the FI $z=0$;  i.e., we assume that a magnon that approaches the 
surface from below with momentum $(\bm{q}_{\parallel},q_z)$ is scattered by the interface into 
a state with momentum $(\bm{q}_{\parallel},-q_z)$ [Fig. \ref{fig2}(a)]. 
This approximation neglects diffuse scattering effects due to surface 
roughness and does not account for interactions between the magnon system and the TI surface quasiparticles.
We show later that the presence of the TI has a negligible influence on the FI magnon distribution.
Equation (\ref{steadymagboltz}) is an inhomogeneous first-order linear differential equation which we solve by 
integrating backward along the path followed by the magnon to reach a given position.  
For magnons at position $z <0 $ that have positive (toward the surface) group velocity, this 
path does not include reflection, whereas for magnons that have negative (away from the surface) group velocity,
the path includes specular reflection at an earlier time.  In this way we obtain that for $q_z\geq0$
\begin{align}
& \delta n_{\bm{q}_{\parallel},q_z}(z)  =-\tau_m |v_{q_z}|\partial_zT\frac{\partial n^{(0)}_{\bm{q}}}{\partial T},\notag\\
& \delta n_{\bm{q}_{\parallel},-q_z}(z) =\tau_m |v_{q_z}|\partial_zT\frac{\partial n^{(0)}_{\bm{q}}}{\partial T}\left[1-2\exp\left(-\frac{|z|}{|v_{q_z}|\tau_m} \right)\right].
\label{deln}
\end{align}
Note that 
\begin{align}
n_{\bm{q}_{\parallel},q_z}(z=0)=n_{\bm{q}_{\parallel},-q_z}(z=0).
\label{specular}
\end{align}

Far from the surface, the temperature gradient induces a magnon current, but because of 
cancellation between $q_z > 0$ and $q_z < 0$ response, it does not change the magnon density.  
Close to the surface, the cancellation is imperfect.  
Using Eq. ($\ref{deln}$), we obtain an expression for the 
total nonequilibrium magnon density: 
\begin{align}
\delta n(z)&\equiv\int_{q_z\geq0}\frac{d^3q}{(2\pi)^3}[\delta n_{\bm{q}_{\parallel},q_z}+\delta n_{\bm{q}_{\parallel},-q_z}]\notag\\
&\sim -\frac{\tau_m\partial_zT}{\pi^2}\int_0^1 dt\int_0^{q_T} dq\ qt  \exp\left(-\frac{|z|}{2D\tau_mqt } \right).\label{totalaccum}
\end{align}
In the second approximate version of this integrand, we have set $\partial n^{(0)}_{\bm{q}}/\partial T \to 1/\omega_{\bm{q}}$ for 
$|\bm{q}|\leq q_T\equiv \sqrt{T/D}$, and set it to zero for $|\bm{q}|>q_T$.  Although this approximate expression 
can be integrated analytically, the result is not particularly transparent.  We have plotted the total 
nonequilibrium magnon distribution obtained by accurately integrating 
Eq. (\ref{totalaccum}) in Fig. \ref{fig2}(b) where we see that a nonequilibrium magnon population builds up 
at the surface, where $l_m\equiv2Dq_T\tau_m$ is a characteristic magnon mean free path.
In the following sections, we consider the interaction between the nonequilibrium magnons accumulated at the
interface and the electrons on the TI surface states.  We see
that the spin-Seebeck voltage signal depends not only on the total nonequilibrium magnon 
density, but also on its momentum distribution in relation to the Fermi surface of 
the TI surface states.

\section{Topological-insulator Dirac cone response to nonequilibrium magnons\label{modelsec}}
In this section, we formulate a semiclassical theory of the TI's Dirac cone surface-state response to 
nonequilibrium magnons.  Using an electron Boltzmann equation with an electron-magnon-scattering collision
term, we are able to obtain an expression of the electric field generated in the TI surface-state 
system by the temperature gradient across the FI.  

\subsection{Model of the interface}
We model\cite{hasan,xlq} the TI surface states by a spin-momentum-locked 
Dirac Hamiltonian:
\begin{align}
H_e&= A \int\frac{d^2k}{(2\pi)^2}\psi_{\bm{k}}^{\dagger}\hat{\mathcal{H}}_e(\bm{k})\psi_{\bm{k}},\notag\\
\hat{\mathcal{H}}_e(\bm{k})&=vk_x\hat{\sigma}_y-vk_y\hat{\sigma}_x-\mu \hat{1}\notag\\
&=\sum_{\alpha=\pm}\xi^{\alpha}_{\bm{k}}|\bm{k},\alpha \; \rangle\langle \bm{k},\alpha|,\label{electronham}
\end{align}
where $A$ is the system area, 
($\psi,\psi^\dagger$) are two-component creation and annihilation 
spinors for the surface-state electrons, $\bm{k}=(k_x,k_y)$ is the two-dimensional electron 
momentum, $v$ is the Fermi velocity, $\mu$ is the chemical potential, and 
$\hat{\sigma}_i$ are Pauli matrices that act in spin space.
In the second line, we define projection 
operators $|\bm{k},\pm\rangle\langle \bm{k},\pm|=(\hat{1}\pm\bm{d}(\bm{k})\cdot\hat{\bm{\sigma}})/2$
for the upper and lower Dirac bands with energies $\xi^{\pm}_{\bm{k}}=\pm v|\bm{k}|-\mu$.
Here $\bm{d}(\bm{k})=(-\sin \theta_{\bm{k}},\cos\theta_{\bm{k}},0)$, and $\theta_{\bm{k}}$ is the 
momentum $\bm{k}$ orientation angle.  

We assume that the surface-state quasiparticles are exchange-coupled to the surface magnetization of 
the TI: 
\begin{align}
 H_{exc}=-\frac{J a }{2}\int d^3x\delta(z)\psi^\dagger(x,y)\hat{\bm{\sigma}}\psi(x,y)\cdot\bm{S}(\bm{x}),
\end{align}
where $a$ is the lattice constant of the FI, and $J$ characterizes the strength of the exchange coupling.  The mean-field coupling between the TI 
quasiparticles and the $y$ direction ground-state magnetization yields only an irrelevant
shift in the $k_x$ direction in momentum space, which has no consequence.  
For small fluctuations in magnetization direction, the remaining interaction Hamiltonian 
can be rewritten as an electron-magnon interaction:
\begin{align}
H_{em}&=-\frac{J aA^3}{2}\sum_{i=x,z}\int\frac{d^2kd^2q_{\parallel}}{(2\pi)^2(2\pi)^2} \;  \psi^\dagger_{\bm{k}}\hat{\sigma}_i\psi_{\bm{k}+\bm{q}_{\parallel}}S^i_{\bm{q}_{\parallel}}(z=0)\notag\\
&=g\frac{A^3}{a^2}\int\frac{d^2kd^2q_{\parallel}}{(2\pi)^2(2\pi)^2} \psi^\dagger_{\bm{k}}\hat{\sigma}^+\psi_{\bm{k}+\bm{q}_{\parallel}}a^{\dagger}_{\bm{q}_{\parallel}}(z=0)+\mathrm{H.c.},\label{emham}
\end{align}
where $\bm{q}_{\parallel}$ is the in-plane component of the magnon momentum, $g=-\sqrt{2S_0}J/4$, and 
$\hat{\sigma}^\pm=\hat{\sigma}_z\pm i\hat{\sigma}_x$.

\subsection{Electron Boltzmann equation}
We concentrate on physics near the Fermi surface and ignore interband scattering.
As mentioned in Ref. \onlinecite{jiang} and explicitly proven in the Appendix,
the spin-Seebeck electric field is invariant under a particle-hole transformation 
$\mu\rightarrow-\mu$ in the simple Dirac model (see  for the exact proof). 
In the following, we therefore assume that $\mu>0$, drop band indices 
and include only the conduction 
band ($|\bm{k} \rangle\equiv|\bm{k},+ \rangle$, and $\xi_{\bm{k}}\equiv \xi^+_{\bm{k}}$), 
and measure momenta 
relative to the new Dirac point after the shift produced by the interaction with the 
ground state magnetization has been applied.

\begin{figure}[t]
\begin{center}
\includegraphics[width=8cm,angle=0,clip]{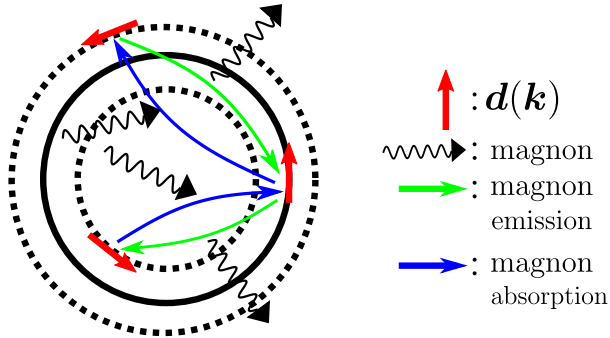}
\caption{Schematic illustration of magnon emission and absorption. 
The solid and dotted circles are constant-energy surfaces for the conduction band and the 
arrows indicate the importance of spin-momentum locking in the Dirac cone.
Magnon emission lowers energy and is accompanied by a 
$\hat{\sigma}^+$ electronic operator, whereas magnon absorption increases energy 
and is accompanied by a $\hat{\sigma}^-$ electronic operator.}
\label{fig3}
\end{center}
\end{figure}

To describe the topological SSE,
we consider the linearized Boltzmann equation:
\begin{align}
\frac{\partial f_{\bm{k}}}{\partial t}-e\bm{E}^{em}\cdot\bm{v}_{\bm{k}}\frac{\partial f^{(0)}_{\bm{k}}}{\partial \xi_{\bm{k}}}=\left.\frac{\partial f_{\bm{k}}}{\partial t}\right|_{imp}+\left.\frac{\partial f_{\bm{k}}}{\partial t}\right|_{em},\label{eleboltzmann}
\end{align}
where $\bm{E}^{em}$ is the induced electric field, $\bm{v}_{\bm{k}}=(v_x,v_y)=v(\cos\theta_{\bm{k}},\sin\theta_{\bm{k}})$, $f_{\bm{k}}$ is the momentum-dependent electron distribution function, and $f^{(0)}_{\bm{k}}=(\exp(\xi_{\bm{k}}/T) +1)^{-1}$ is the Fermi distribution function at temperature $T$.
The terms on the right-hand side are the electron-impurity and electron-magnon-scattering 
collision terms, respectively.
The electron-magnon-scattering term can be calculated by using the quantum Fokker-Planck equation,\cite{dwsnoke}
and is given to second order in the electron-magnon interaction by
\begin{widetext}
\begin{align}
& \left.\frac{\partial  f_{\bm{k}}}{\partial t}\right|_{em}=2\pi g^2a^3\int\frac{d^2q_{\parallel}dq_z}{(2\pi)^3}  \notag\\ 
& \times\Big[  \; |\langle \bm{k}+\bm{q}_{\parallel}|\hat{\sigma}^-|\bm{k} \rangle|^2 \; 
\delta (\omega_{\bm{q}}+\xi_{\bm{k}}-\xi_{\bm{k}+\bm{q}_{\parallel}}) 
\left[(1-f_{\bm{k}})f_{\bm{k}+\bm{q}_{\parallel}}(1+n_{\bm{q}_{\parallel},q_z}(z=0))-f_{\bm{k}}(1-f_{\bm{k}+\bm{q}_{\parallel}})n_{\bm{q}_{\parallel},q_z}(z=0)              \right]\notag\\
+\; &|\langle \bm{k}-\bm{q}_{\parallel}|\hat{\sigma}^+|\bm{k} \rangle|^2
\delta (\omega_{\bm{q}}-\xi_{\bm{k}}+\xi_{\bm{k}-\bm{q}_{\parallel}})
\left[(1-f_{\bm{k}})f_{\bm{k}-\bm{q}_{\parallel}}n_{\bm{q}_{\parallel},q_z}(z=0)    -f_{\bm{k}}(1-f_{\bm{k}-\bm{q}_{\parallel}})(1+n_{\bm{q}_{\parallel},q_z}(z=0))             \right] \Big] , 
\label{qfp}
\end{align}
\end{widetext}
where the $|\langle \bm{k'}|\hat{\sigma}^{\pm}|\bm{k} \rangle|^2$ factors
account for the influence of spin-momentum locking in the Dirac cone on the electronic transition probabilities 
associated with magnon emission and absorption (Fig. \ref{fig3}).
These electronic matrix elements 
can be calculated by observing that the projection 
operator $|\bm{k}\rangle\langle \bm{k}|=(\hat{1}+\bm{d}(\bm{k})\cdot\hat{\bm{\sigma}})/2$:

\begin{align}
|\langle \bm{k'}|\hat{\sigma}^{\pm}|\bm{k} \rangle|^2&=\mathrm{Tr}\left[\hat{\sigma}^{\pm}|\bm{k}\rangle\langle \bm{k}|\hat{\sigma}^{\mp}|\bm{k'}\rangle\langle \bm{k'}|\right]\notag\\
&=(1\mp d^y(\bm{k}))(1\pm d^y(\bm{k'})).\label{probability}
\end{align}

The nonequilibrium nature of the SSE is captured by
the statistical factors in small square brackets in Eq. ($\ref{qfp}$), which can be further simplified.
In linear response we can replace $f_{\bm{k}}$ by $f^{(0)}_{\bm{k}}$ 
in the electron-magnon-scattering term. 
Since the right-hand side of Eq. ($\ref{qfp}$) is zero by detailed balance, when
the magnons are also in equilibrium, 
the two square-bracket statistical factors can be replaced by the following two factors:
\begin{align}
&(f^{(0)}_{\bm{k}+\bm{q}_{\parallel}}-f^{(0)}_{\bm{k}})\delta n_{\bm{q}_{\parallel},q_z}(z=0),\notag\\
&(f^{(0)}_{\bm{k}-\bm{q}_{\parallel}}-f^{(0)}_{\bm{k}})\delta n_{\bm{q}_{\parallel},q_z}(z=0).
\end{align}
With this replacement the electron-magnon-scattering term is replaced explicitly 
in terms of the nonequilibrium correction to the magnon distribution function 
at the interface $\delta n_{\bm{q}_{\parallel},q_z}(z=0)$.  Next, the integral over $q_z$ 
in Eq. ($\ref{deln}$) can be evaluated using the energy conservation $\delta$ functions, 
\begin{align}
\int_{-\infty}^{\infty} dq_z |v_{q_z}|\delta(\omega_{\bm{q}}-X)&=2\int_{0}^{\infty}dq_z (2Dq_z)\delta(\omega_{\bm{q}}-X),\notag\\
&=2\Theta(X-D|\bm{q}_{\parallel}|^2),
\end{align}
to obtain 
\begin{widetext}
\begin{align}
\left.\frac{\partial  f_{\bm{k}}}{\partial t}\right|_{em}=g^2a^3(-2\tau_m\partial_zT)\int\frac{d^2q_{\parallel}}{(2\pi)^2}&\Big[ |\langle \bm{k}+\bm{q}_{\parallel}|\hat{\sigma}^-|\bm{k} \rangle|^2
\Theta (\xi_{\bm{k}+\bm{q}_{\parallel}}-\xi_{\bm{k}}-D|\bm{q}_{\parallel}|^2)
(f^{(0)}_{\bm{k}+\bm{q}_{\parallel}}-f^{(0)}_{\bm{k}})\frac{\partial n^{(0)}(\xi_{\bm{k}+\bm{q}_{\parallel}}-\xi_{\bm{k}})}{\partial T}  \notag\\
+&|\langle \bm{k}-\bm{q}_{\parallel}|\hat{\sigma}^+|\bm{k} \rangle|^2
\Theta (\xi_{\bm{k}}-\xi_{\bm{k}-\bm{q}_{\parallel}}-D|\bm{q}_{\parallel}|^2)
(f^{(0)}_{\bm{k}-\bm{q}_{\parallel}}-f^{(0)}_{\bm{k}})\frac{\partial n^{(0)}(\xi_{\bm{k}}-\xi_{\bm{k}-\bm{q}_{\parallel}})}{\partial T} \Big] .\label{qfpeq}
\end{align}
\end{widetext}
Note that the free integral over $q_z$, present because the three-dimensional magnon system 
is driving a two-dimensional electronic system, replaces the usual Fermi-golden-rule $\delta$ function
by a step function.

Since  we do not assume any particular property of the 
TI surface states in Eq. ($\ref{qfpeq}$), 
our theory applies to any single-band two-dimensional electron system that is exchange-coupled
to the surface magnetism of a ferromagnetic insulator, and is 
simply generalized to multiband two-dimensional materials by adding band indices.

\begin{figure*}[t]
\begin{center}
\includegraphics[width=14cm,angle=0,clip]{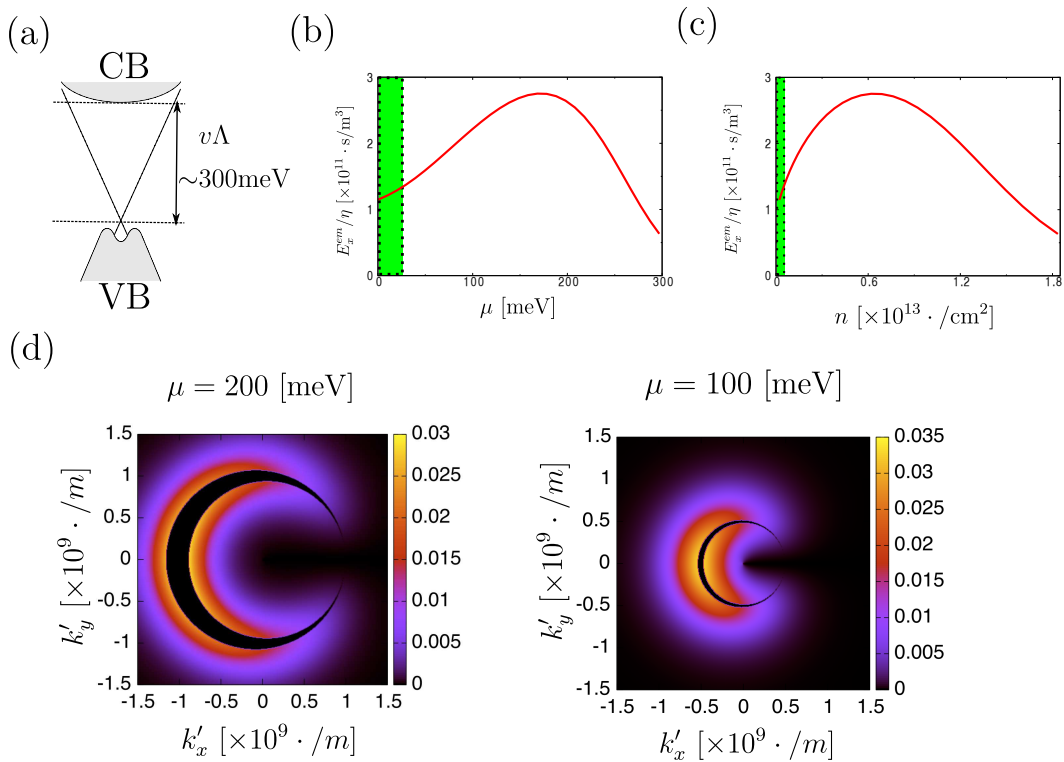}
\caption{(a) Schematic illustration of the band structure of a Bi$_2$Te$_3$ film.
The shaded regions labeled VB and CB are the bulk valence and conduction bands, respectively. 
The surface-state Dirac point is much closer to the valence band than to the conduction band.
The thermally electric field at $T=300$ K is plotted in (b) vs chemical potential and in (c) vs electron density.
In the green region ($\mu\leq T$), the results are not accurate since we neglect the interband effect.
(d) The integrand of Eq. ($\ref{qfpeq}$) in arbitrary units for $\bm{k}=(k_F,0)$ 
as a function of $\bm{k'}$ for chemical potential $\mu=$100 and 200 meV relative to the 
Dirac point.  The electron-magnon interaction vertex tends to be strongest for transitions between
electronic states with opposite momentum.}
\label{fig4}
\end{center}
\end{figure*}

\subsection{Induced electric field in the steady state}
We are now in a position to derive an expression
for the electric field induced by the electron-magnon interaction in the steady state.
The nonequilibrium transport of the topological-insulator surface state can be also described by the Boltzmann equation \cite{culcer}.
For simplicity, we use a relaxation-time approximation for the electron-impurity 
collision term in the steady-state electron Boltzmann equation:
\begin{align}
-e\bm{E}^{em}\cdot \bm{v}_{\bm{k}}\frac{\partial f^{(0)}_{\bm{k}}}{\partial \xi_{\bm{k}}}=-\frac{\delta f_{\bm{k}}}{\tau_e}+\left.\frac{\partial f_{\bm{k}}}{\partial t}\right|_{em},\label{eleboltzmann}
\end{align}
where $\delta f_{\bm{k}}=f_{\bm{k}}-f^{(0)}_{\bm{k}}$, and $\tau_e$ is the relaxation time.
Since the spin-Seebeck voltage is measured under open circuit conditions, it can be evaluated by 
finding the electric field strength at which the electric current vanishes: 
\begin{align}
\int\frac{d^2k}{(2\pi)^2}\bm{v}_{\bm{k}}\delta f_{\bm{k}}=0.\label{netcurrent}
\end{align}
Using Eqs. ($\ref{eleboltzmann}$) and ($\ref{netcurrent}$), we find that 
\begin{align}
E_i^{em}=\left.\left[\int \frac{d^2k}{(2\pi)^2} \, v_i \left.\frac{\partial f_{\bm{k}}}{\partial t}\right|_{em}\right]\right/\left[-e\int \frac{d^2k}{(2\pi)^2} v_i^2\frac{\partial f^{(0)}_{\bm{k}}}{\partial \xi_{\bm{k}}}\right].\label{defofe}
\end{align}
In deriving Eq. (\ref{defofe}), we have appealed to isotropy in asserting that
$\int d^2k v_xv_y=0$.  Note that $E_i^{em}$ is independent of the electron-disorder scattering time 
$\tau_{e}$.  

In the topological SSE for magnetization in the 
$y$ direction, the induced electric field is in the $x$-direction: $\bm{E}^{em}=(E^{em}_x,0)$.
To see this, we rewrite Eq. ($\ref{probability}$) as
\begin{align}
|\langle \bm{k'}|\hat{\sigma}^{\pm}|\bm{k} \rangle|^2=&\pm\cos\theta_{\bm{k}}(\cos \Delta\theta_{\bm{k},\bm{k'}}-1)\notag\\
&-\sin\theta_{\bm{k}}(\cos\theta_{\bm{k}}\mp1)\sin\Delta\theta_{\bm{k},\bm{k'}}\notag\\
&+(1-\cos^2\theta_{\bm{k}}\cos \Delta\theta_{\bm{k},\bm{k'}}),\label{angledep}
\end{align}
where $\Delta\theta_{\bm{k},\bm{k'}}\equiv \theta_{\bm{k}}-\theta_{\bm{k'}}$.
The second line on the right-hand side of Eq. (\ref{angledep}) does not contribute to 
Eq. ($\ref{qfpeq}$) because the other factors in the integrand are even functions of $\Delta\theta_{\bm{k},\bm{k'}}$. 
The term on the third line does not contribute to the numerator in Eq. ($\ref{defofe}$) because its contribution to
the integrand is odd in $\sin\theta_{\bm{k}}$ or $\cos\theta_{\bm{k}}$.
Similarly the first line contributes  to $E^{em}_x$ ($v_x \propto \cos\theta_{\bm{k}}$), but 
not to $E^{em}_y$ ($v_y \propto \sin\theta_{\bm{k}}$).   
The electric-field direction predicted by our theory is 
consistent with the experimental result \cite{jiang}.

In the following sections, we retain only the first line on the right-hand side of Eq. ($\ref{angledep}$) since the other 
terms do not contribute to the final result.  
For notational convenience, we therefore rewrite Eq. ($\ref{qfpeq}$) as
\begin{align}
&\left.\frac{\partial \delta f_{\bm{k}}}{\partial t}\right|_{em}=g^2a^3(-2\tau_m\partial_zT)\cos\theta_{\bm{k}}A(vk),\label{asymmetric}
\end{align}
where $A$ is a $\theta_{\bm{k}}$-independent function.

\section{Numerical results\label{numerics}}
To compare our theory with experiment\cite{jiang}, we compute the integrals in the 
numerator and denominator of Eq. (\ref{defofe}).  
For numerical estimates, we use $v=3.0\times10^5$ m/s 
for the surface state Dirac velocity of Bi$_2$Te$_3$\cite{Jzhang} and 
$D=5.0\times10^{-21}$ eVm$^2$ for the spin stiffness of YIG\cite{shinozaki,srivastava}.
The momentum cutoff $\Lambda$ of the TI surface states is fixed by setting 
$v \Lambda$ equal to the bulk band gap\cite{Jzhang} $\sim 300$ meV,
which yields $\Lambda \sim1.5\times10^9$m$^{-1}$
[see Fig. $\ref{fig4}$(a)].
The spin-Seebeck signal is proportional to $\eta \equiv 2g^2a^3\tau_m\partial_zT/e$.

\subsection{Trend versus carrier density}
Since the motivating experiment is performed at room temperature, we first set $T=300$ K and 
examine in Fig. $\ref{fig4}$(b) the dependence of the spin-Seebeck signal on the position of the Fermi level.  
We find a nonmonotonic dependence that is illustrated 
by plotting the thermally induced electric field $E^{em}_x$
as a function of the chemical potential $\mu$ in Fig. $\ref{fig4}$(b)
and as a function of electron density $n\equiv \int d^2k/(2\pi)^2 f^{(0)}_{\bm{k}}$ in Fig. $\ref{fig4}$(c).
The nonmonotonic behavior arises from  the momentum distribution of the nonequilibrium magnons
combined with the increase with $\mu$ in the electronic density of states, 
which enhances the phase space available for electron-magnon scattering.
To illustrate this anomalous behavior, we plot in Fig. \ref{fig4}(d) the integrand of Eq. ($\ref{qfpeq}$)
for fixed $\bm{k}=(k_F,0)$, where $k_F=\mu/v$ is the Fermi momentum, 
as a function of $\bm{k'}=\bm{k}\pm\bm{q}$.
Because the electron-magnon interaction vertex reverses spins relative to the 
magnetization direction, the integrand tends to be stronger for transitions between
electronic states with opposite momentum directions.  
At the larger chemical potential ($\mu=200$ meV) value, 
large angle scatterings are suppressed due to the lack of available magnons, 
which is captured by the step functions and the other part of the integrand. 
Suppression by the step function is weaker at smaller 
chemical potential ($\mu=100$ meV). 
Our use of a momentum cutoff, which crudely captures the less stark 
spin-momentum coupling in bulk and higher-energy surface-state bands,
also competes with the density-of-states effect to produce a maximum spin-Seebeck 
signal at a finite chemical potential.  Because we neglect the role of valence-band 
states, our calculations are not accurate at very small carrier densities ($\mu\lesssim T$).
In our calculations $E^{em}_x$ has its maximum value
at $n=6\times10^{12}\ \mathrm{cm}^{-2}$.
In experiment, the electric field at 
$n=4\times10^{12}\ \mathrm{cm}^{-2}$ is found to be $\sim 50$ times greater than that at 
$n=2\times10^{13}\ \mathrm{cm}^{-2}$.
Our numerical result explains the electric field enhancement at relatively small 
electron densities, although our simplified model does not achieve quantitative agreement.

\begin{figure}[]
\begin{center}
\includegraphics[width=7cm,angle=0,clip]{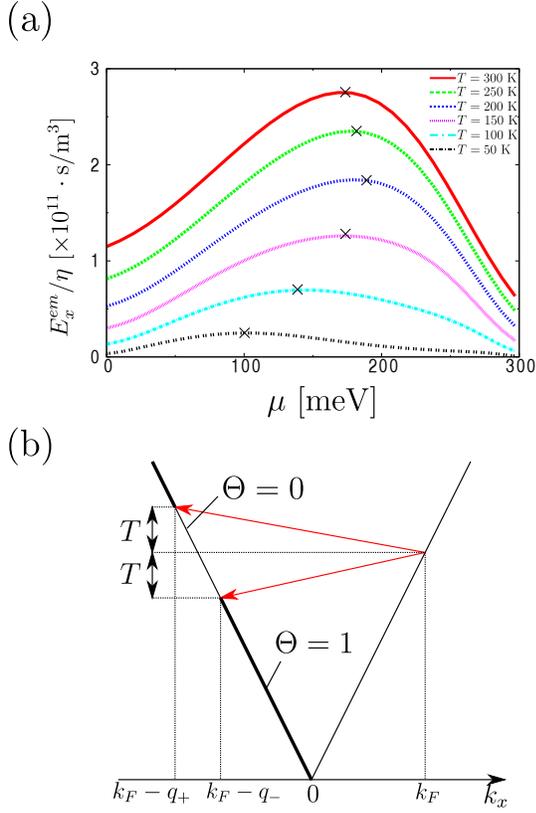}
\caption{(a) Spin-Seebeck electric field $E^{em}_x$ as a function of $\mu$ for various temperatures. 
The cross marks denote the largest electric field points for each temperature. (b) Schematic illustration
of electron-magnon backscattering $\bm{k}=(k_F,0)\rightarrow\bm{k'}=(-|\bm{k'}|,0)$ at $\mu=\mu^{est}_{max}$.}
\label{fig5}
\end{center}
\end{figure}

\subsection{Order-of-magnitude estimate\label{orderestimation}}
In order to estimate the numerical size of the spin-Seebeck signal at room temperature,
we use the following typical values for YIG\cite{bauer}: $S_0 \sim10$ and $a\sim1\times10^{-9} {\rm m}$.
Assuming that the dominant magnon relaxation mechanism is related to 
Gilbert damping of the macroscopic magnetization at 
room temperature (see Sec. \ref{magdiscussion}), we 
use $\tau_m\sim\hbar/\alpha_G k_BT\sim1\times10^{-9}$ s\cite{gilbert,bender,bauer},
where $\alpha_G\sim1\times10^{-4}$.
Finally we set the interface exchange coupling to $J \sim1$ meV, 
the same order as for a Pt/YIG interface.
From Fig. \ref{fig4}(b), the maximum value of $E^{em}_x/\eta $ at room temperature is 
$\sim3\times 10^{11}$ s/m$^3$.
Using the above values to estimate $\eta$, 
we obtain $E^{em}_x\sim0.9$ V/m for $\partial_z T\sim5\times10^3$ K/m, which is of the 
same order of magnitude as the experimental value of $\sim0.2$ V/m.

\subsection{Temperature dependence of induced electric field}
The $\mu$ dependence of $E^{em}_x$ is plotted for various temperatures in Fig. \ref{fig5}(a).
The overall trend is that $E^{em}_x$ increases with 
temperature because of the increase in magnon population.
The chemical potential at which $E^{em}_x$ reaches its maximum, 
$\mu_{max}$, decreases with decreasing temperature for $T\lesssim200$ K.
$\mu_{max}$ can be roughly estimated by considering 
backscattering contributions, e.g., $\bm{k}=(k_F,0)\rightarrow\bm{k'}=(-|\bm{k'}|,0)$.
For each $k_F$, we define $q_{\pm}$ such that 
\begin{align}
Dq_{\pm}^2=\pm\left(v(q_{\pm}-k_F)-vk_F\right).
\end{align}
Because of the magnon contribution to the final-state energy, 
the step functions in the integrand of Eq. (16) both vanish 
for $q_--k_F<|\bm{k'}|<q_+-k_F$ [see Fig. \ref{fig5}(b)].
For $|q_{\pm}-2k_F|\ll k_F$, $q_{\pm}\simeq2k_F\pm 4Dk_F^2/v$.
In this approximation, the step functions are zero for
\begin{align}
k_F-\frac{4Dk_F^2}{v}\lesssim |\bm{k'}|\lesssim k_F+\frac{4Dk_F^2}{v}.
\end{align}
Since the magnon statistical factors have large values for $k_F-T/v\lesssim |\bm{k'}|\lesssim k_F+T/v$,
the overlap with the step functions is large when $k_F<\sqrt{T/4D}$ .
Roughly speaking, $\mu_{max}$ is expected to be given by
\begin{align}
\mu_{max}^{est}=\frac{v}{2}\sqrt{\frac{T}{D}}.\label{muest}
\end{align}
According to Eq. ($\ref{muest}$), $\mu^{set}_{max}=$100, 140, 170, 200, 220, and 240 meV for $T=$50, 100, 150, 200, 250, and 300 K, respectively, in good agreement with the more accurate 
values in Fig. $\ref{fig5}$(a) for $T\lesssim200$ K.
Equation ($\ref{muest}$) cannot explain the maximum values for $T=$250 and 300 K, where $\mu^{est}_{max}$ is close to $v\Lambda\sim300$ meV, because these considerations do not account for the Dirac surface state cutoff.

\section{discussions\label{discussion}}
\subsection{The magnon collision integral\label{magdiscussion}}

We now return to discuss our use of a relaxation time approximation in the 
magnon Boltzmann equation  [Eq. (\ref{magrelaxterm})].
At room temperature, magnon-phonon-scattering processes dominate relaxation of the magnon distribution function.
Our assumption of a characteristic time over which any nonequilibrium
magnon population will approach equilibrium when undriven, is consistent with stochastic Landau-Lifshitz-Gilbert 
equations\cite{hoffman} in which magnetization relaxation appears in the 
Gilbert damping term (see Sec. \ref{orderestimation}).
It is nevertheless important to distinguish the roles of processes that conserve magnon number,
for example, processes in which a magnon and a phonon exchange energy and momentum,
from processes that do not conserve magnon number, for example, ones in which magnons are 
converted into phonons and vice versa.  If the former processes are strongly dominant, 
a possibility proposed by Cornelissen $et$ $al.$ \cite{bauer}, the non-equilibrium magnon 
distribution can assume a different form characterized by a local chemical potential.
The change in distribution might have a quantitative influence on the relationship between the 
excess magnon population at the magnetic-insulator/topological-insulator interface, and hence 
the topological SEE, and the 
Gilbert damping parameter used for our quantitative estimates.  
The magnon-phonon conversion processes that we have in mind in using the 
relaxation time approximation, correspond to changes in the ground-state 
magnetization configuration, in response to changes in the lattice.  
The momentum dependence of the relaxation time for these processes was investigated in Ref. \onlinecite{spinwave}.  Although we ignored momentum dependence in our simple relaxation-time 
approximation, its inclusion would not change our results drastically 
because the dominant processes for the spin-Seebeck effect have $|\bm{q}|\sim q_T$.

A less important issue in our theory is our assumption of
specular reflection.  Because the magnetic-insulator/topological-insulator interface is 
often rough, it would be more realistic to use boundary conditions that allow for a mixture of 
specular and diffuse boundary scattering.  We make the simpler assumption mainly to avoid 
introducing another model parameter whose value is not accurately known.  
Because the excess magnon population, $\delta n_{\bm{q}_{\parallel},q_z}$, 
does not depend on the 
direction of $\bm{q}_{\parallel}$, we do not expect much of an influence of diffuse 
surface scattering. 

We have also assumed that the influence of electron-magnon scattering at the surface on the 
magnon distribution function is negligible.
To justify this approximation, we now compare the rate of change of  
total spin due to magnon-electron scattering with the total magnon relaxation rate.
The spin injection rate is given by
\begin{align}
\left.\frac{\partial s^y}{\partial t}\right|_{em}\equiv \int \frac{d^2k}{(2\pi)^2}\frac{d^{y}(\bm{k})}{2}\left.\frac{\partial f_{\bm{k}}}{\partial t}\right|_{em}.\label{spininjection}
\end{align}
For an order-of-magnitude estimate, we use the following approximations:
\begin{align}
&\Theta (|\xi_{\bm{k'}}-\xi_{\bm{k}}|-D|\bm{k}-\bm{k'}|^2)\sim1,\label{app1}\\
&k_F-\frac{T}{v}\lesssim |\bm{k'}|\lesssim k_F+\frac{T}{v},\label{app2}\\
&(f^{(0)}_{\bm{k'}}-f^{(0)}_{\bm{k}})\frac{\partial n^{(0)}(\xi_{\bm{k'}}-\xi_{\bm{k}})}{\partial T}\sim \frac{\partial f^{(0)}_{\bm{k}}}{\partial \xi_{\bm{k}}}.\label{app3}
\end{align}
Using Eqs. (\ref{app1}), (\ref{app2}), (\ref{app3}), and $\partial f^{(0)}_{\bm{k}}/\partial \xi_{\bm{k}}\sim -\delta(\xi_{\bm{k}})$, we can rewrite Eqs. (\ref{qfpeq}) and (\ref{spininjection}) as
\begin{align}
\left.\frac{\partial f_{\bm{k}}}{\partial t}\right|_{em}\sim
&\frac{8\pi g^2a^3k_F(\tau_m T\partial_zT)}{v}\delta(\xi_{\bm{k}})\cos\theta_{\bm{k}},\\
\left.\frac{\partial s^y}{\partial t}\right|_{em}\sim&\frac{g^2a^3k_F^2(\tau_mT\partial_zT)}{v^2}.
\end{align}

In comparison, the total magnon relaxation rate is given by
\begin{align}
&\int_{-\infty}^{0} dz\frac{\delta n(z)}{\tau_m}\notag\\
&\sim -\frac{\partial_zT}{\pi^2}\int_{-\infty}^{0} dz\int_0^1 dt\int_0^{q_T} dq\ qt  \exp\left(-\frac{|z|}{2D\tau_mqt }  \right)\notag\\
&=-\frac{2q_T(\tau_mT\partial_zT)}{9\pi^2},
\end{align}
where we have used Eq. (\ref{totalaccum}) in the second line.
Thus, the ratio of the spin injection rate to the total magnon relaxation rate is approximately given by
\begin{align}
\left. \left.\frac{\partial s^y}{\partial t}\right|_{em}\right/\left|\int_{-\infty}^{0} dz\frac{\delta n(z)}{\tau_m}\right|\sim\frac{9\pi^2g^2a^3k_F^2}{2v^2q_T}.\label{ratio}
\end{align}
The value of the right hand side of 
Eq. (\ref{ratio}) is $\sim1\times10^{-3}$ for $k_F,q_T\sim1\times10^{9}$m$^{-1}$.
It follows that the influence of the TI on the magnon distribution in the FI 
is indeed negligible.
Note that the ratio does not depend on the relaxation time $\tau_m$ or, equivalently, the Gilbert damping constant $\alpha_G$.

\subsection{Corrections to our simplified model}
Although our theory explains the topological SSE 
qualitatively, estimating the correct order of magnitude of the effect and
its carrier-density dependence,  it does overestimate $E^{em}_x$ in comparison to 
experiment, especially for large $\mu$.
There are mainly two possible reasons.
First, both the magnitude and surface-state energy dependence of the 
exchange coupling $J$ is uncertain.
Second, the surface state in experimentally realized topological insulators is not 
very accurately described by the simple Dirac electron model.
The leading correction\cite{lianfu} in Bi$_2$Te$_3$ is a hexagonal warping\cite{ylchen,dhsieh} correction:
\begin{align}
\hat{\mathcal{H}}(\bm{k})\propto (k_x^3-3k_xk_y^2)\hat{\sigma}_z.\label{warping}
\end{align}
This term implies that $\bm{d}(\bm{k})$ has a large out-of-plane component for large chemical potentials,
which would have the effect of reducing the SSE strength.

\subsection{Thermal gradient generation in the $x$ direction}
\begin{figure}[]
\begin{center}
\includegraphics[width=7cm,angle=0,clip]{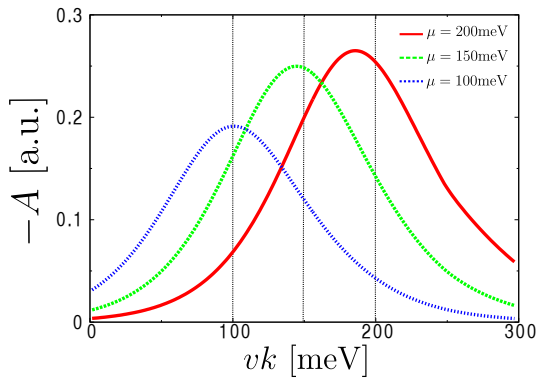}
\caption{Energy dependence of the factor  
$A$ which captures the energy dependence of the 
Dirac state population response at a given orientation,
as a function of $vk$ for several $\mu$ values.
These results were calculated at room temperature.}
\label{fig6}
\end{center}
\end{figure}

Since the origin of the electron-magnon scattering term is a temperature gradient,
we expect that energy currents will also flow in the topological insulator.
In Fig. \ref{fig6}, we plot $A(vk)$ for various chemical 
potentials.  Because $A(vk)$ is an asymmetric function with respect to $vk=\mu$,
it follows that electron-magnon scattering drives not only
electrical signal but also heat flow.  We therefore predict that the $\partial_z T$, 
applied temperature gradient of the experimental spin-Seebeck geometry, will 
induce an $x$-direction temperature gradient $\partial_xT$.

\subsection{SSE in a magnetically doped TI}
\begin{figure}[]
\begin{center}
\includegraphics[width=8cm,angle=0,clip]{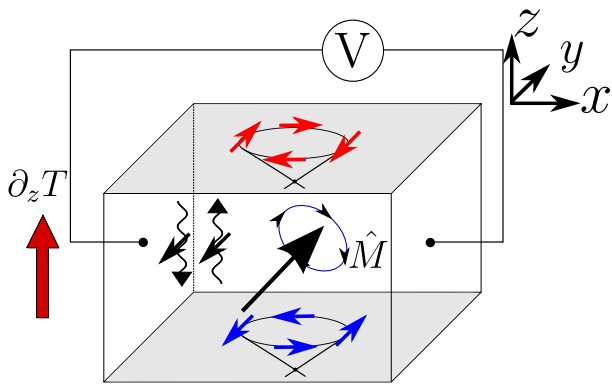}
\caption{Schematic illustration of the SSE in a magnetically doped TI thin film.}
\label{fig7}
\end{center}
\end{figure}

We now generalize our discussion to magnetically doped TIs 
like Cr-doped (Bi Sb)$_2$Te$_3$\cite{kou1,czchang1,czchang2,fan}, 
Cr-doped Bi$_2$Se$_3$\cite{kou2}, and Mn-doped Bi$_2$(Te Se)$_3$\cite{checkelsky,ylchen2}.
The anomalous Hall effect in these systems demonstrates robust magnetic order,
at least at very low temperatures.  Under these conditions 
a magnetic TI under a vertical temperature gradient (Fig. \ref{fig7}) can be 
described by a model similar to that discussed elsewhere in this paper.
If the film is uniformly magnetized, the two surface-state systems will be coupled to the 
same magnon gas.
Because the effective Hamiltonians of the top and bottom surface states have different chiralities,
\begin{align}
\hat{\mathcal{H}}^{top(bottom)}(\bm{k})=\pm\left[-vk_x\hat{\sigma}_y+vk_y\hat{\sigma}_x\right]-\mu \hat{1},
\end{align}
where top (bottom) corresponds to $+$ ($-$),
the signs of the magnon accumulation at two surfaces are opposite, 
and the charge currents generated at top and bottom surfaces will have the
same sign.  We therefore predict that a transverse voltage will be 
induced by a vertical temperature gradient in magnetically doped TI thin films.

\begin{acknowledgments}

The authors acknowledge helpful interactions with
Z. Jiang, C. Z. Chang, J. S. Moodera and  J. Shi.
This work was performed as part of the SHINES, an Energy Frontier Research Center funded by the U.S. Department of Energy, Office of Science, Basic Energy Sciences under Award  \#SC0012670. 
N. O. is supported by the Japan Society for the Promotion of Science (JSPS) through the Program for Leading Graduate Schools (MERIT). N. O. is also supported by JSPS KAKENHI (Grant No. 16J07110).

\end{acknowledgments}
\appendix 
\section{Dirac-spinor particle-hole transformation\label{phsym}}

We prove here that for an ideal symmetric Dirac cone, the thermally induced electric field is 
unchanged under $\mu\rightarrow-\mu$. 
For convenience, we introduce the hole Dirac 
spinors $(\phi,\phi^{\dagger})$ defined in terms of $(\psi,\psi^{\dagger})$ as
\begin{align}
\phi_{\bm{k}}=\hat{\sigma}_y\psi^{\dagger}_{-\bm{k}},\notag\\
\phi^{\dagger}_{\bm{k}}=\psi_{-\bm{k}}\hat{\sigma}_y,
\end{align}
or, equivalently,
\begin{align}
\phi(\bm{x})=\hat{\sigma}_y\psi^{\dagger}(\bm{x}),\notag\\
\phi^{\dagger}(\bm{x})=\psi(\bm{x})\hat{\sigma}_y.
\end{align}
This particle-hole transformation is defined such that the matrix representation of the spin density operator is unchanged:
\begin{align}
\bm{s}(x)\equiv\psi^{\dagger}(\bm{x})\left[\frac{\hat{\bm{\sigma}}}{2}\right]\psi(\bm{x})=\phi^{\dagger}(\bm{x})\left[\frac{\hat{\bm{\sigma}}}{2}\right]\phi(\bm{x}).
\end{align}
The electron Dirac Hamiltonian (\ref{electronham}) can be rewritten in terms of $(\phi,\phi^{\dagger})$ as
\begin{align}
H_e&=\int\frac{d^2k}{(2\pi)^2}\psi_{\bm{k}}^{\dagger}\hat{\mathcal{H}}_e(\bm{k})\psi_{\bm{k}}\notag\\
&=(\mathrm{const})+\int \frac{d^2k}{(2\pi)^2}\psi_{-\bm{k},i}[-\hat{\mathcal{H}}^*_e(-\bm{k})]_{i,j}\psi^{\dagger}_{-\bm{k},j}\notag\\
&=(\mathrm{conts})+\int\frac{d^2k}{(2\pi)^2}\phi_{\bm{k}}^{\dagger}[-\hat{\sigma}_y\hat{\mathcal{H}}^*_e(-\bm{k})\hat{\sigma}_y]\phi_{\bm{k}}\notag\\
&=(\mathrm{const})+\int\frac{d^2k}{(2\pi)^2}\phi_{\bm{k}}^{\dagger}[-(vk_x\hat{\sigma}_y-vk_y\hat{\sigma}_x)+\mu \hat{1}]\phi_{\bm{k}}.\label{holeham}
\end{align}
Equation (\ref{holeham}) shows that the particle-hole transformation changes the sign of $\mu$ and the chirality of the Dirac Hamiltonian.
The form of the electron-magnon Hamiltonian ($\ref{emham}$), on the other hand, does not change under the particle-hole transformation:
\begin{align}
H_{em}=g\int\frac{d^2kd^2q_{\parallel}}{(2\pi)^2(2\pi)^2} \phi^\dagger_{\bm{k}}\hat{\sigma}^+\phi_{\bm{k}+\bm{q}_{\parallel}}a^{\dagger}_{\bm{q}_{\parallel}}(z=0)+\mathrm{H.c.},
\end{align}
since the magnon operators couple not to the charge density but to the hole spin density that has the same matrix representation as the electron spin density.

In the following, we consider the electron Dirac Hamiltonian with $\mu=-|\mu_0|$.
This Hamiltonian is equivalent to the hole Dirac Hamiltonian with $\mu=|\mu_0|$.
The scattering term ($\ref{qfpeq}$) for this hole Dirac Hamiltonian has the same magnitude but the opposite sign as that for the electron Dirac Hamiltonian with $\mu=|\mu_0|$ due to its opposite chirality, while the definition of the electric field ($\ref{defofe}$) for the hole Dirac Hamiltonian has the opposite sign as that for the electron Dirac Hamiltonian due to its opposite charge.
Thus, the induced electric field for the hole Dirac Hamiltonian with $\mu=|\mu_0|$ is the same as that for the electron Dirac Hamiltonian with $\mu=|\mu_0|$ or, equivalently, the induced electric field  for the electron Dirac Hamiltonian is unchanged as $\mu_0\rightarrow-\mu_0$.

\end{document}